# *NIRVANA* AI GOVERNANCE: HOW AI POLICYMAKING IS COMMITTING THREE OLD FALLACIES

Jiawei Zhang*

In his well-known article, *Information and Efficiency: Another Viewpoint*, Harold Demsetz points out a common fantasy in public policy called the *Nirvana* approach.[1] Under the *Nirvana* approach, policymakers "seek to discover discrepancies between the ideal and the real and if discrepancies are found, they deduce that the real is inefficient."[2] He argues that the *Nirvana* approach involves three logical fallacies, namely, the "grass is always greener on the other side" fallacy, the "free lunch" fallacy, and the "people could be different" fallacy.[3] My research finds that various AI governance proposals also adopt the *Nirvana* approach with these three inherent fallacies.

## I. "*THE GRASS IS ALWAYS GREENER ON THE OTHER SIDE*" FALLACY

Harold Demsetz quotes Kenneth J. Arrow's arguments that "a free enterprise economy [will] underinvest in invention and research (as compared with an ideal) because it is risky … [and therefore] for optimal allocation to invention, it would be necessary for the government or some other agency not governed by profit-and-loss criteria to finance research and invention."[4] Demsetz then criticized it as a "the grass is always greener on the other side" fallacy.[5] Demsetz critically notes that Arrow does not fully explain why the counterparts—here, the government and nonprofit agency—can perform a better job than a free enterprise solution. "Whether the free enterprise solution can be improved upon by the substitution of the government or other nonprofit

* Lloyd M. Robbins Doctor of Juridical Science (J.S.D.) Fellow at U.C. Berkeley Law School; Ph.D. Researcher at the Chair for Public Policy, Governance and Innovative Technology at the Technical University of Munich.

[1] *See* Harold Demsetz, *Information and Efficiency: Another Viewpoint*, 12 J.L. & ECON. 1 (1969).

[2] *Id.* at 1.

[3] *Id.* at 2.

[4] *Id.*

[5] *Id.*

institutions in the financing of research cannot be ascertained *solely by examining the free enterprise solution*."[6] It is apparently unfair to say that the grass on the other side of the fence is greener just because the grass on this side is not that green. Demsetz thus provides a warning to policymakers that the substantive differences between the existing situation and proposed replacements should be sufficiently examined before implementing that proposal.

"The grass is always greener on the other side" fallacy is very common in AI governance. This fallacy has two manifestations. First, some commentators believe that people are more reliable than machines. They identify the weaknesses of machines without, however, examining the limitations of human beings. This is fallacious because it is not enough to argue for human oversight just because AI is not fully reliable. It is still necessary to examine whether human involvement can realize a better outcome.[7] This fallacy derives largely from a cognitive bias, which is referred to as "algorithm aversion"—people would rather accept inferior human decisions than superior algorithmic decisions.[8] Cass R. Sunstein and Jared H. Gaffe identify several factors that account for the algorithm aversion, such as:

1. a desire for agency;
2. a negative moral or emotional reaction to judgment by algorithms;
3. a belief that certain human experts have unique knowledge, unlikely to be held or used by algorithms;
4. ignorance about why algorithms perform well; and
5. asymmetrical forgiveness, or a larger negative reaction to algorithmic error than to human error.[9]

The error of overreliance on human decision-making has been well documented in occasional failures of what is known as Reinforcement Learning from Human Feedback (RLHF).[10] RLHF was originally designed to draw on human efforts to fine-tune AI models to achieve better alignment with societal expectations.[11] However, it is not necessarily effective as the accuracy of human feedback is also heavily subject to participants' subjectiveness.[12] Excessive human intervention can even backfire. Google's AI chatbot "Gemini," which was artificially trained to be more "accurate" and "inclusive," ended up generating historically inaccurate images, such as a woman as pope, black Vikings, female National Hockey League players, and a black George Washington.[13]

---

[6] *Id.* (emphasis added).

[7] For an excellent comparison between human decisions and AI decisions and criticisms of the assumption that human decisions are inherently superior, see generally Cary Coglianese & Alicia Lai, *Algorithm vs. Algorithm*, 72 DUKE L.J. 1281 (2021); see also Ben Green, *The Flaws of Policies Requiring Human Oversight of Government Algorithms*, 45 COMPUT. L. & SEC. REV. 105681 (2022) (finding that "people are able to effectively oversee algorithmic decision-making").

[8] *See* Berkeley J. Dietvorst, Joseph P. Simmons & Cade Massey, *Algorithm Aversion: People Erroneously Avoid Algorithms After Seeing Them Err*, 144 J. EXPERIMENTAL PSYCH. 114, 114 (2015); Cass R. Sunstein & Jared H. Gaffe, *An Anatomy of Algorithm Aversion*, 26 COLUM. SCI. & TECH. L. REV. 290, 290 (2025); Viktor Mayer-Schönberger, *Of Forests and Trees in AI Governance*, J. CYBER POL'Y 1, 1 (2025).

[9] Sunstein & Gaffe, *supra* note 8, at 290, 300–08.

[10] *See, e.g.*, Ryan Lowe & Jan Leike, Aligning Language Models to Follow Instructions, OPENAI (Jan. 27, 2022), https://openai.com/index/instruction-following/ [https://perma.cc/V2HV-HMCA].

Second, some believe that the government works better in controlling AI risks than companies' self-regulation. They see the profit-driven nature of private AI companies but, intentionally or not, neglect the government's inadequacies in various aspects relative to private entities.[14] For example, some researchers argue for ex ante licensure, regulatory sandbox, and AI auditing to enhance government involvement in the development process of AI models. However, it is insufficient to only present the AI risks and the incapabilities of AI companies; it is still necessary to justify the role of the government by explaining why the government, compared to private AI companies, is more capable of handling the present issues under the licensure, regulatory sandbox, and auditing regimes. But the government is not necessarily well-positioned to mitigate AI risks given its lack of technical expertise, inadequate information capture and processing, and untimely and unagile response to the changing situation.[15] In fact, AI companies' self-regulation sometimes proves effective, especially when improvement of their services is visible to general users and aligned with public expectations.[16] AI companies, under sufficient market forces, will keep upgrading their AI model performance and derisking their systems automatically, voluntarily, and continuously.[17] Policymakers must realize the inherent values of the market mechanism and potential government failures when proposing to regulate AI with heightened government interference.

---

[11] *See generally* Long Ouyang et al., Training Language Models to Follow Instructions with Human Feedback (Mar. 4, 2022) (unpublished manuscript) (on file with arXiv), https://arxiv.org/pdf/2203.02155 [https://perma.cc/UKB8-VC7A].

[12] *See, e.g.*, *id.* at 19 ("Some of the labeling tasks rely on value judgments that may be impacted by the identity of our contractors, their beliefs, cultural backgrounds, and personal history."); *see also* Luke Munn, Liam Magee & Vanicka Arora, *Truth Machines: Synthesizing Veracity in AI Language Models*, 39 AI & SOC'Y 2759, 2763 (2023) ("[RLHF] is attended by all-too-human subjectivity.").

[13] *See* Thomas Barrabi, *'Absurdly Woke': Google's AI Chatbot Spits Out 'Diverse' Images of Founding Fathers, Popes, Vikings*, N.Y. POST (Feb. 21, 2024), https://nypost.com/2024/02/21/business/googles-ai-chatbot-gemini-makes-diverse-images-of-founding-fathers-popes-and-vikings-so-woke-its-unusable/ [https://perma.cc/2FPR-5H6U].

[14] Regarding government failures, see generally Cass R. Sunstein, *Paradoxes of the Regulatory State*, 57 U. CHI. L. REV. 407 (1990).

[15] *See* Neel Guha et al., *AI Regulation Has Its Own Alignment Problem: The Technical and Institutional Feasibility of Disclosure, Registration, Licensing, and Auditing*, 92 GEO. WASH. L. REV. 1473 (2024); *see also* Jiawei Zhang, *Regulating Chatbot Output via Inter-Informational Competition*, 22 NW. J. TECH. & INTELL. PROP. 109 (2024) (arguing that the market may work more effectively than the government in the context of chatbot output regulation).

[16] *See* MAURICE E. STUCKE & ARIEL EZRACHI, COMPETITION OVERDOSE: HOW FREE MARKET MYTHOLOGY TRANSFORMED US FROM CITIZEN KINGS TO MARKET SERVANTS chs. 1&2 (2020).

[17] *See, e.g.*, *id.*; Zhang, *supra* note 15, at 135.

## II. "*Free Lunch*" Fallacy

Demsetz argues that Arrow's commodity-option proposal slipped into the fallacy of the free lunch, as Arrow does not sufficiently consider the cost of the commodity option.[18] "The cost of marketing commodity options exceed[ing] the gain from the adjustment to risk," Demsetz explains, "would account for their presumed absence" of the commodity options.[19] He emphasizes the role of scarcity in evaluating the real-world problem and generating a policy proposal.[20] Policymakers should realize that just because the reality does not match the ideal does not mean it is nonoptimal. Comparing the real world with inevitable scarcity to an ideal one without considering the costs of achieving that goal is unrealistic and misleading.

Some AI policy proposals also commit the "free lunch" fallacy. Policymakers and researchers sometimes overlook that when they come up with novel regulatory solutions targeting a specific AI problem, the harms and costs are also inherent in their proposals.[21] If the proposer does not weigh the costs against the benefits of their proposal and examine other comparable alternatives that can achieve similar regulatory objectives, their proposal will be less convincing. One example is Article 4 of the Chinese Generative AI Interim Measure.[22] This provision adopts a zero-risk standard by listing exhaustive illegal possibilities.[23] The policymakers did not realize that this standard, which is so high as to be unattainable, would cause significant chilling effects on generative AI service providers, converting ChatGPT into a "Sorry"GPT. The resulting costs are reduced welfare for public consumers, as they cannot access their desired response from AI chatbots.

## III. "*The People Could Be Different*" Fallacy

Demsetz challenges Arrow's contention that moral hazard constitutes "a unique and irremediable cause of incomplete coverage of all risky activities by insurance."[24] Instead, Demsetz argues that "the moral hazard problem is no different than the problem posed by any cost."[25] Moreover, "some risks are left uninsured because the cost of moral hazard is too great and this may mean that self-insurance is economic."[26] He identifies this fallacy as the idea that *if people were different*—such as not engaging in moral hazards—then the real world would be more efficient or at least more in line with theoretical ideals. This fallacy overlooks the fact that people's subjective preferences, limitations, and imperfect behavior are part of reality.[27] Therefore, policymaking should not be based on a comparison between the real world, where people act imperfectly, to a hypothetical, idealized world where people behave perfectly. Instead, policy decisions should be designed based on how they perform given the real nature of human behavior.

[18] Demsetz, *supra* note 1, at 2–4.

[19] *Id.* at 4.

[20] *See id.*

[21] *See generally* Guha et al., *supra* note 15.

[22] *See* Shengcheng Shi Rengong Zhineng Fuwu Guanli Zanxing Banfa (生成式人工智能服务管理暂行办法) [Interim Measures for the Administration of Regulating Generative AI Services] (promulgated by Cyberspace Admin., Nat'l Dev. & Reform Comm'n, Ministry Educ., Ministry Sci. & Tech., Ministry Indus. & Info. Tech., Ministry Pub. Sec., Nat'l Radio & Television Admin.), July 10, 2023, http://www.cac.gov.cn/2023-07/13/c_1690898327029107.htm [https://perma.cc/YL7X-QRDR]. For an English-translated version, see, for example, Interim Measures for the Management of Generative Artificial Intelligence Services, CHINA L. TRANSLATE (July 13, 2023), https://wwwchinalawtranslate.com/en/generative-ai-interim/ [https://perma.cc/94GH-682C].

[23] *Id.* at art. 4.

This fallacy also manifests in AI governance. When designing AI regulatory tools and setting standards, some researchers and policymakers are prone to impose more and more harsh and even zero-risk approaches. This tendency is derived from the false comparison between *the AI-driven world where AI does lead to some risks* and *an entirely idealized world where no risk exists at all*. This fallacy has fueled many unrealistic proposals to govern AI risks. For instance, some researchers propose imposing "truth-telling" duties on large language models to cure their "careless speech."[28] Such proposals are established on the assumption that AI speech should be aligned with a parallel world where people never produce careless speech. This is unrealistic. In fact, large language models merely learn real-world problems from their input and replicate them in their predictions.[29] "The more frequently a claim appears in the dataset, the higher the likelihood it will be repeated as an answer."[30]

Similarly, some policymakers and researchers embrace explainability rules to enhance the transparency of the AI decision-making process.[31] Undeniably, AI explainability requirements can help clarify and justify AI-made decisions, but they should not be set on an unrealistic standard. John Zerilli and others have perceptively found that some policies are adopting a double standard where "machine tools must be transparent to a degree that is in some cases unattainable, in order to be considered transparent at all, while human decision-making can get by with reasons satisfying the comparatively undemanding standards of practical reason."[32] Actually, the human brain, like AI, also acts like a black box—it can be inherently biased, but the workings of the brain are non-understandable and non-detectable[33]—and humans are skilled at using beautiful language to decorate their decisions and gloss over their true reasons.[34] However, we never require human decisions to be fully explainable as we require AI.

---

[24] Demsetz, *supra* note 1, at 7.

[25] *Id.*

[26] *Id.*

[27] *See id.*

[28] *See generally* Sandra Wachter, Brent Mittelstadt & Chris Russell, *Do Large Language Models Have a Legal Duty to Tell the Truth?*, 11 ROYAL SOC'Y OPEN SCI. 240197 (2024), https://royalsocietypublishing.org/doi/10.1098/rsos.240197 [https://perma.cc/8FZK-76Y8].

[29] *See* Jiawei Zhang, *ChatGPT as the Marketplace of Ideas: Should Truth-Seeking Be the Goal of AI Content Governance?*, 35 STAN. L. & POL'Y REV. ONLINE 11, 28–29 (2024), https://law.stanford.edu/publications/comment-chatgpt-as-the-marketplace-of-ideas/ [https://perma.cc/669P-6EEY]; Tonja Jacobi & Matthew Sag, *We Are the AI Problem*, 74 EMORY L.J. ONLINE 1 (2024), https://scholarlycommons.law.emory.edu/cgi/viewcontent.cgi?article=1049&context=elj-online [https://perma.cc/63NQ-X5A5].

[30] Munn, Magee & Arora, *supra* note 12, at 2761.

[31] *See generally* Sandra Wachter, Brent Mittelstadt & Luciano Floridi, *Transparent, Explainable, and Accountable AI for Robotics*, 2 SCI. ROBOTICS 1 (2017).

The appropriate approach is to compare the AI-driven world to *a real world where risks are everywhere and where people are not perfectly rational but can live well with these risks*. Policymakers should understand that our world inherently involves various risks; some AI problems are merely a part of societal risks, be it bias and discrimination, misinformation, data disclosure, environmental footprints, lack of accountability, or opaque decision-making processes. It is fallacious to say that we must cleanse AI risks just because AI *has* some risks. We must identify how AI *enhances or magnifies* the risks which have long existed in our real world before raising regulatory proposals.

## IV. CONCLUSION

This essay has applied Harold Demsetz's concept of the *Nirvana* approach to the realm of AI governance and illuminated three common fallacies in various AI policy proposals: the "grass is always greener on the other side" fallacy, the "free lunch" fallacy, and the "people could be different" fallacy. By doing so, I have exposed fundamental flaws in how policymakers and researchers often approach AI governance. The prevalence of these fallacies in AI governance underscores a broader issue: the tendency to idealize potential solutions without fully considering their real-world implications. This idealization can lead to regulatory proposals that are not only impractical but potentially harmful to innovation and societal progress.

However, this research does not challenge any specific proposal conclusion but rather critiques the underlying mindsets and logical frameworks that inform these proposals. This research serves as a critical reminder that effective AI governance requires a nuanced, comparative approach. Researchers and policymakers when generating a regulatory proposal should (1) rigorously compare proposed alternatives and the status quo, considering the strengths and weaknesses of both; (2) acknowledge that there is no "free lunch" in policy implementation and carefully weigh the costs against the benefits of new regulatory measures; and (3) base standard-setting and policy design on realistic expectations of human and AI behavior rather than unattainable ideals. Following these suggestions will enable us to craft a more balanced, pragmatic, and effective framework for AI governance.

---

[32] John Zerilli et al., *Transparency in Algorithmic and Human Decision-Making: Is There a Double Standard?*, 32 PHIL. & TECH. 661, 668 (2019).

[33] *Id.* at 674–75; SCOTT PLOUS, UNDERSTANDING PREJUDICE AND DISCRIMINATION 17 (2003).

[34] *See* Zerilli et al., *supra* note 32, at 675.